\documentclass[
  ,draft            
  ]
  {aipproc}

\layoutstyle{6x9}

\begin{document}

\title{Dynamo Action in Fully Convective Low-Mass Stars}

\classification{95.30.Lz, 95.30.Qd, 97.10.Cv, 97.10.Jb, 97.10.Kc, 97.10.Ld,
97.20.Jg, 97.20.Vs}
\keywords      {convection -- MHD -- stars: magnetic fields -- turbulence}

\author{Matthew K. Browning}{
  address={Astronomy Dept, UC Berkeley, 601 Campbell Hall, Berkeley CA 94720-3411}
}

\author{Gibor Basri}{
  address={Astronomy Dept, UC Berkeley, 601 Campbell Hall, Berkeley CA 94720-3411}
}

\begin{abstract}
  Recent observations indicate that fully convective stars can effectively
  build magnetic fields without the aid of a tachocline of shear, that
  those fields can possess large-scale components, and that they may sense
  the effects of rotation.  Motivated by these puzzles, we present global
  three-dimensional simulations of convection and dynamo action in the
  interiors of fully convective M-dwarfs of 0.3 solar masses.  We use the
  Anelastic Spherical Harmonic (ASH) code, adopting a spherical
  computational domain that extends from 0.08-0.96 times the overall
  stellar radius.  We find that such fully convective stars can generate
  magnetic fields of several kG strength, roughly in equipartition with the
  convective flows.  Differential rotation is established in hydrodynamic
  progenitor calculations, but essentially eliminated in MHD simulations
  because of strong Maxwell stresses exerted by the magnetic fields.
  Despite the absence of interior angular velocity contrasts, the magnetic
  fields possess strong mean (axisymmetric) components, which we attribute
  partly to the very strong influence of rotation upon the slowly
  overturning flows.  

\end{abstract}

\maketitle


\section{MOTIVATING QUESTIONS}

Of all the many findings that have emerged from solar physics over the last
two decades, the discovery of the tachocline of shear likely figures as the
most significant for dynamo theory.  Revealed by helioseismology (e.g.,
Gough \& Toomre 1991), the tachocline is a narrow boundary layer situated
near the base of the convection zone, in which the solar angular velocity
transitions from differential rotation in the unstable envelope to nearly
solid-body rotation beneath it.  Its discovery motivated the now-prevalent
``interface dynamo'' paradigm, in which the tachocline plays a pivotal
role -- both because of its shear, which helps stretch and organize
toroidal fields, and because of the stable stratification, which may allow
fields to be greatly amplified before becoming susceptible to magnetic
buoyancy (e.g., Ossendrijver 2003).   Recently, simulations have helped affirm the likely
importance of the tachocline in generating the Sun's organized magnetism:
global simulations that include a tachocline (Browning et al. 2006;
Browning et al. 2007) give rise to magnetic fields that appear more
``solar-like'' than do simulations of the convection zone alone (Brun,
Miesch \& Toomre 2004).  (See proceedings by Miesch et al. 2007, this
volume.)  And if the tachocline is vital in building the Sun's orderly
magnetism, then similar boundary layers are probably likewise central in
the dynamo action of any star that possesses both a convective envelope and
a radiative, stable core.

But not all stars are so configured.  Moving down the main sequence to
stars of lower mass, the size of the inner stable layer gradually
decreases.  By around spectral class M3, corresponding to a mass of about
0.35 solar masses, it has vanished entirely, and convection occurs
throughout the interior.  In a fully convective star, there is no
radiative-convective interface, and so presumably no interface dynamo --
implying that if fields are built in such stars, the dynamo responsible
must operate rather differently than in the Sun.  A natural conclusion is
that such fully convective stars should show surface magnetic field
behavior quite distinct from that realized in their more massive cousins
(e.g., Durney, De Young \& Roxburgh 1993; Chabrier \& Kuker 2006).

Surprisingly, observations of low-mass stars have partly confounded even
this rather basic expectation.  In a recent breakthrough, Donati et
al. (2006) used Zeeman Doppler imaging to constrain the surface magnetic
field morphology of a very rapidly rotating fully convective M-dwarf.  The
kG-strength fields they detected were predominantly axisymmetric and
large-scale, in contrast to the purely small-scale fields predicted by some
models (e.g., Durney et al. 1993).  Donati et al. (2006) also found that
their target star did not exhibit any surface differential rotation.
Separately, other observations have demonstrated that strong magnetic
activity appears to be quite common in fully convective stars: indeed, the
fraction of M-stars showing with detectable activity actually peaks at
about spectral class M8 (West et al. 2004).

Further puzzles have come from study of the remarkable relation between
stellar magnetic activity and rotation rate.  In stars like the Sun,
observations indicate that chromospheric and coronal activity increase
with rotation rate, then ``saturate'' above a threshold velocity (e.g.,
Noyes et al. 1984; Delfosse et al. 1998; Pizzolato et al. 2003).  Many
authors have noted that this rotation-activity correlation is tightened
considerably when rotation is expressed in terms of the Rossby number,
estimated in these studies as just $P_{\rm rot}/\tau_{c}$, with $\tau_c$ a
typical convective overturning time; likewise, the threshold rotational
period needed for ``saturated'' activity is apparently a function of
stellar mass, varying from about a day in early-G stars to about ten days
in early M-stars (Pizzolato et al. 2003).  This rotation-activity
connection appears to persist in some fashion into the mid-M stars: Mohanty
\& Basri (2003) argued that a sample of stars ranging from M0 to M5 showed
a common ``saturation-type''  rotation-activity relationship, with observed
activity roughly independent of rotation rate above a threshold value.
Because measuring the rotation rates of the slowest rotators is difficult,
it remains unclear whether magnetic activity in these stars increases
gradually with rotation rate as in solar-like stars, or instead transitions
more abruptly.  In the late-M spectral classes and beyond, the
rotation-activity relation does appear to break down (e.g., Reiners et
al. 2007), but this may occur well beyond the point where stars become
fully convective.  In all, it appears that fully convective stars can
certainly act as magnetic dynamos, can (in at least some cases) build
large-scale magnetic fields (Donati et al. 2006), and may sense the effects
of rotation in roughly the same fashion as more massive stars.

Motivated by these puzzles, we have carried out global 3--D nonlinear
simulations of convection and dynamo action in the interiors of fully
convective M-dwarfs.  In the sections that follow, we briefly describe our
modeling (\S2), its principal results (\S3), and the implications of our
work (\S4).  The simulations here are described in more detail in Browning
(2007).

\section{MODEL FORMULATION}

Our simulations are intended to be simplified models of 0.3 solar mass
M-stars rotating at the solar angular velocity ($\Omega = 2.6 \times
10^{-6}$ s$^{-1}$).  We, like a host of others at this meeting, utilize the
Anelastic Spherical Harmonic (ASH) code, which solves the 3--D
Navier-Stokes equations with magnetism in the anelastic approximation
(Clune et al. 1999; Miesch et al. 2000; Brun, Miesch \& Toomre 2004).  Our
spherical computational domain extends from 0.08-0.96R, with R the overall
stellar radius of $2.07 \times 10^{10}$ cm.  We exclude the inner few
percent of the star from our calculations both because the coordinate
systems employed in ASH are singular there, and because the small numerical
mesh sizes at the center of the star would require impractically small
timesteps.  Our computations also do not extend
all the way to the stellar surface, because the very low densities in the
outer few percent of the star favor the driving of fast, small-scale
motions that we cannot resolve.

The initial stratifications of the mean density, energy generation rate,
gravity, radiative diffusivity, and entropy gradient $dS/dr$ are adopted
from a 1-D stellar model (I. Baraffe, private communication, after Baraffe
\& Chabrier 2003).  We update these thermodynamic quantities throughout the
course of the simulation, as the evolving convection modifies the
spherically symmetric mean state.  Variables are expanded in terms of
spherical harmonic basis functions $Y_l^m(\theta, \phi)$ in the horizontal
directions and Chebyshev polynomials $T_n(r)$ in the radial.  In the
simulations here, we have retained spherical harmonic degrees up to
$\ell_{\rm max} =340$ (implying $N_{\theta}=512$ and $N_{\phi}=1024$) and
retain $N_r=192$ Chebyshev colocation points.  As with all numerical
simulations, we must employ eddy viscosities and diffusivities that are
vastly greater than their counterparts in actual stars; here we have taken
these to be constant in radius, and adopted a Prandtl number
$\nu/\kappa=0.25$ and a magnetic Prandtl number $Pm=\nu/\eta=8$.  We also
conducted other simulations at varying $\nu$,$\eta$, and $Pm$, but have
chosen to focus on one case in this paper for clarity; the others are
briefly described in Browning (2007).

Only one prior large-scale numerical MHD simulation has modeled fully
convective stars (Dobler, Stix \& Brandenburg 2006, hereafter DSB06).  Our
approach differs from theirs in a few key ways.  In order to keep thermal
relaxation timescales small, DSB06 rescaled the stellar luminosity to a
value many orders of magnitude greater than appropriate for an actual M
dwarf; because this implies a commensurate increase in typical convective
velocities, they also were forced to consider very rapid rotation rates in
order to keep the Rossby number of their simulations realistic.  This
rescaling is needed in their simulations partly because they adopt the same
thermal diffusivities for the mean temperature gradient and for the
small-scale turbulent temperature fluctuations; in ASH, the mean
temperature gradient is acted on by the thermal diffusivity $\kappa_r$
taken from a 1--D stellar model, whereas $\kappa$ for the turbulent
temperature field is (as in DSB06) a sub-grid-scale eddy diffusivity.  Thus
we employ the actual stellar luminosity, rotation rate, etc. in our
modeling.  The method adopted in ASH allows to examine the radiative flux
in the interior with reasonable fidelity, since $\kappa_r$ in our models is
essentially set by the radiative opacities of the 1--D stellar model.
Neither strategy is perfect: ours does not allow for adjustments to the
thermal stratification that occur over the very long thermal timescale.
Our simulations also differ from those of DSB06 in a few smaller ways.  The
overall density contrast between the inner and outer boundaries in our
models is about 170, consistent with the contrast between 0.1 and 0.96R in
the 1--D stellar model we used for our initial conditions.  In DSB06, the
density varied by a factor of about 5 from center to surface; the larger
density contrasts in our modeling have substantial impact on the morphology
of the convective flows.  The boundary conditions adopted in DSB06 also
differ from ours; theirs is closer to a no-slip boundary condition than to
the stress-free boundaries used here.

\section{FLOWS AND MAGNETIC FIELDS REALIZED}

\subsection{Morphology of the Flows}

The convective flows realized in our simulation possess structure on many
scales.  Near the surface, there is a marked asymmetry between upflows and
downflows: the former tend to be broad and weak, whereas the latter are
strong and narrow.  This asymmetry is a generic feature of turbulent
compressible convection (e.g., Brummell et al. 2002), and arises mainly
because of the strong density stratification.  At depth, the flows are
weaker and of larger physical scale: motions can span large fractions of a
hemisphere and extend radially for great distances.  Motions at the two
depths are linked, with small downflow plumes near the surface coalescing
as they descend to form the broader flows in the deep interior.  The
contrast between flows at the two depths is apparent in Figure 1$a$,$d$,
which show the radial velocity on spherical surfaces at $r=0.88R$ and
$r=0.24R$ respectively.  The flow amplitudes also vary appreciably with
depth, with typical rms velocities declining by a factor of about ten in
going from the surface to the center; because the typical pattern scale of
the convection is also greater at depth, the convective overturning
timescale varies by a factor of about 20 across the domain.

\begin{figure}
  \includegraphics[height=.5\textheight, trim=140 0 140 0]{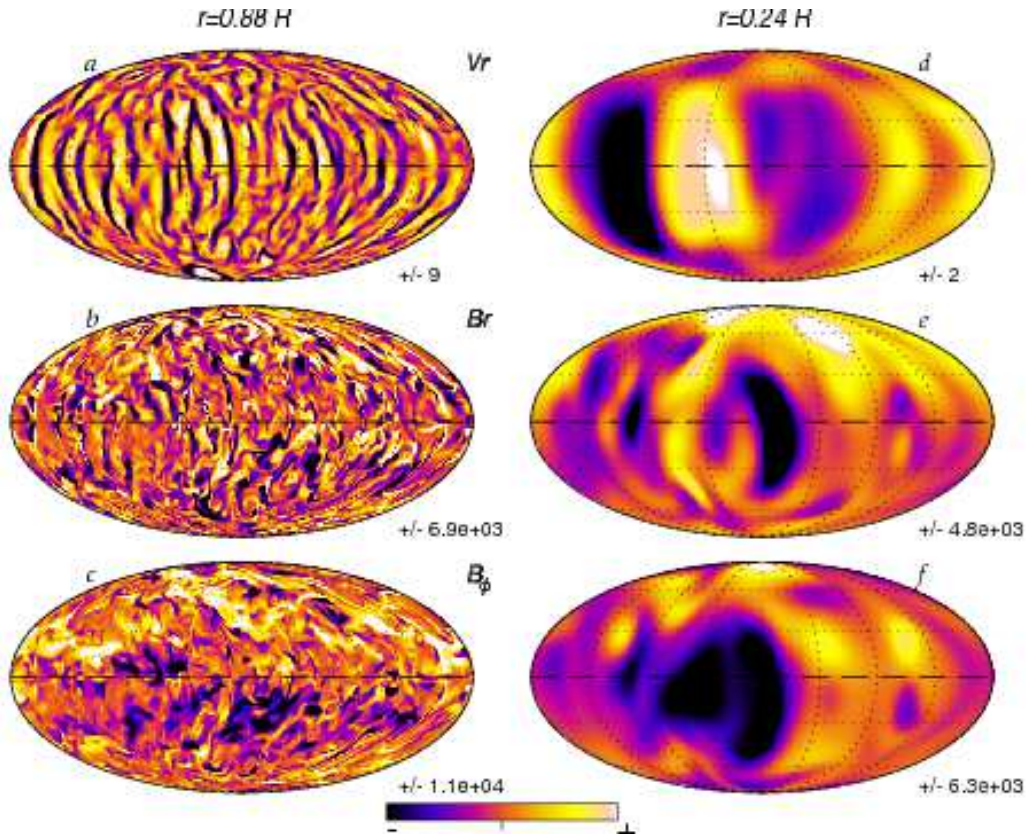}
  \caption{Radial velocity and magnetic fields at a single instant, shown
    on spherical surfaces at two depths.}
\end{figure}

The variation in flow amplitude is linked to both the density
stratification and to variations with radius in the amount of energy that
must be transported by convection.  Although the star is unstably
stratified everywhere, the radiative flux actually carries a majority of
the energy at small radii.  Together with the overall increase of the
total luminosity with radius (out to the radius where nuclear energy
generation stops), this implies that the total luminosity carried by
convection peaks at large radii (around $r=0.80R$). Thus the convective
velocity is also appreciably greater near the surface than at depth.

\subsection{Dynamo Action Achieved}

The flows act as a magnetic dynamo, amplifying a tiny seed field by many
orders of magnitude and sustaining it against Ohmic decay. The magnetic
energy (ME) grows exponentially until it is approximately in equipartition with
the flows.  Over the last 200 days of the simulation, a period during which
no sustained growth or decay of the various energy densities was evident,
ME was approximately 120\% of the total kinetic energy KE (relative to the
rotating frame) and about 140\% of the convective (non-axisymmetric)
kinetic energy (CKE).

As the fields grow, they react back on the flows through the Lorentz force.
Thus KE begins to decline once ME reaches a threshold value of about 5\% of
KE; here this decline is associated mainly with a decrease in the energy of
differential rotation DRKE, whereas CKE remains largely unaffected by the
growing fields.  In the kinematic phase, DRKE is approximately 6 times CKE;
after saturation of the dynamo, DRKE/CKE is only about 0.2-0.4.

\subsection{Morphology of the Fields}

Like the flows that build them, the magnetic fields possess both intricate
small-scale structure as well as substantial large-scale components.  The
typical length scale of the field increases with depth, partly tracking the
radial variation in the size of typical convective flows.  A sampling of
such behavior is provided by Figure 1, which shows the radial field $B_r$
and azimuthal field $B_{\phi}$ on surfaces at two depths.

By decomposing the magnetism into its azimuthal mean (TME), and fluctuations
around that mean (FME), we can gain a coarse estimate of the typical size
of field structures: if the field is predominantly on small scales, only a
small signal will survive this azimuthal averaging.  In these simulations,
TME accounts for about 20\% of the total magnetic energy in the bulk of the
interior; it is smallest near the surface (where TME $\approx$ 5\% ME), and
largest (as a fraction of ME) at depth.

It is striking that the axisymmetric mean fields account for a fairly large
fraction of the total magnetic energy.  In simulations of the bulk of the
solar convective envelope, TME was typically only about 3\% of ME (Brun et
al. 2004); in simulations including a tachocline of shear, similar TME/ME
ratios to those reported here were attained only within the stably
stratified tachocline itself (Browning et al. 2006).  Similarly, Brun,
Browning \& Toomre (2005) found that TME/ME $\approx 0.05$ within most of the convective
cores of A-type stars, with higher values achieved only within a shear
layer at the boundary of that core.  Power spectra constructed for the
simulation here confirm the impressions above: at depth, the magnetic field
is dominated by the largest-scale components, while near the surface it is
more broadly distributed in $\ell$.  Note that the field strength is not
given simply by equipartition at each spherical harmonic degree; rather, ME
can substantially exceed on some scales (see Browning 2007).

The mean (axisymmetric) fields realized in the simulation are remarkably
strong and stable.  Mean toroidal field strengths can exceed 10 kG in some
locations; some prominent field structures persist for thousands of days.
The overall field polarity has flipped only once in roughly 30 years of
simulated evolution: this, too, is in sharp contrast to simulations of
solar convection without a tachocline (Brun et al. 2004), in which the
field polarity flipped at irregular intervals of less than 600 days.

\subsection{Establishment and Quenching of Differential Rotation}

Our hydrodynamical simulation began in a state of uniform rotation, but
convection quickly established interior rotation profiles that varied with
radius and latitude.  The resulting differential rotation, displayed in
Figure 2 (contour plot and panel $b$), was similar to that observed at the solar
surface, in that the equator rotated more rapidly than the poles; unlike
the solar convective envelope, our simulation also exhibited
strong radial angular velocity contrasts, with the outer regions
rotating more rapidly than the interior.  The interior angular velocity
profile was largely constant on cylindrical lines parallel to the rotation
axis, in keeping with the strong Taylor-Proudman constraint.    The overall
angular velocity contrast between the equator and 60 degrees latitude was
about 90 nHz, implying $\Delta \Omega / \Omega \approx$ 22\%.

\begin{figure}
  \includegraphics[height=.3\textheight, trim= 72 36 72 36]{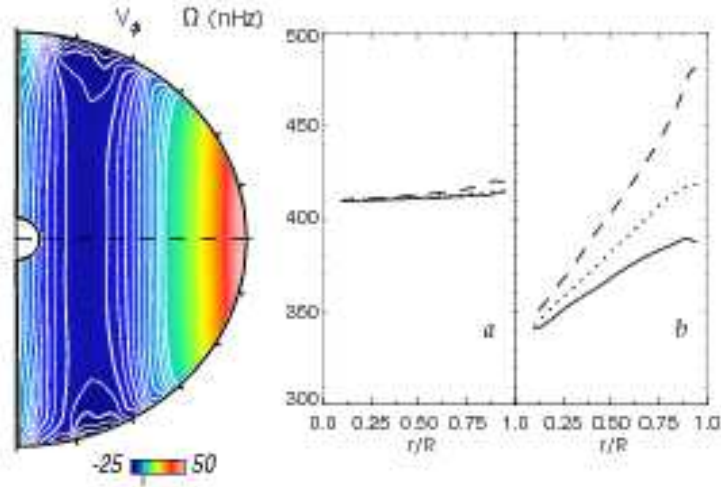}
  \caption{Interior angular velocity established in the hydrodynamic
    progenitor (contour plot and panel $b$), and in the MHD simulation
    (panel $a$).  Shown as a contour plot is the longitudinal velocity
    $V_{\phi}$, averaged in time and in longitude; the right panels show
    the angular velocity $\Omega$ as a function of radius along 
    latitudinal cuts at 0, 30, and 45 degrees.  The MHD simulation rotates
    essentially as a solid body.  }
\end{figure}

The interior rotation profiles are quite different in the presence of
strong dynamo-generated magnetic fields.  In our MHD
simulation, the magnetic fields react back strongly upon the flows, acting
to essentially eliminate the differential rotation.  This behavior is
assessed in Figure 8$a$, which shows the angular velocity
$\hat{\Omega}$ as a function of radius along cuts at various latitudes.
The interior is in nearly solid body rotation; the angular velocity
contrasts realized in the progenitor hydrodynamic case have been almost
entirely eliminated.  This results largely from the action of strong
Maxwell stresses that effectively oppose the transport of angular momentum
by Reynolds stresses.

\section{CONCLUSIONS AND PERSPECTIVES}

The simulation presented here is, we believe, the most faithful description
yet achieved of dynamo action in a fully convective M-dwarf.  Although we
have made many simplifications in our modeling, some of the main findings
may turn out to be robust.  We summarize some of these findings and their
implications here.

The strong stratification plays a major role in setting convective flow
amplitudes and pattern scales, and thus indirectly magnetic field strengths
and morphologies.  Although our model star is unstably stratified
everywhere, there are still two conceptually distinct regions: one near the
surface where convection is quite vigorous and small-scale, and a more
quiescent deep interior.

Convection establishes differential rotation in hydrodynamic cases, but the
angular velocity contrasts are essentially eliminated in MHD simulations.
The strong Maxwell stresses ultimately yield an interior in nearly
solid-body rotation.

The magnetic fields realized here possess structure on many spatial
scales.  The axisymmetric mean component accounts for a surprisingly high
percentage of the total magnetic energy ($\sim$ 20\%), despite the absence
of persistent angular velocity contrasts.  The polarity of mean fields is
remarkably stable, having flipped only once in the $\sim$ 30 years we have
so far evolved the simulation.

We attribute some of these effects to the strong influence of rotation upon
the slowly overturning flows.  Although these simulations rotate at the
solar angular velocity, the influence of rotation is much stronger than in
the Sun because the luminosity, and hence the rms convective velocity, is
much lower.  Comparison to prior simulations of the Sun and A-type stars
(Browning et al. 2004; Brun et al. 2005; Brun et al. 2004; Browning et
al. 2006) suggests that there are plausibly three different regimes of
behavior, separated primarily by the Rossby number.  When rotation is weak,
magnetic energies of less than $\sim$ 30\% KE are typically realized, and
do not greatly modify the differential rotation established in hydrodynamic
cases.  When rotation is somewhat stronger, yielding ME greater than 30\%
KE, cyclical variations of magnetism and differential rotation are
possible; when it is stronger still (as in the simulation here, and the
most rapidly rotating cases of Brun, Browning \& Toomre 2005), ME can
exceed KE without the aid of differential rotation, and any persistent
angular velocity contrasts are strongly quenched.  These conclusions are
also (tentatively) in keeping with calculations of magnetism in rapidly
rotating Sun-like stars (Brown et al., this volume).  These ideas lead to
some straightforward observational predictions -- namely that magnetic,
rapidly rotating M-stars should seldom exhibit differential rotation, that
less rapidly rotating M-stars should generally exhibit weaker fields and
(for very slow rotation) may show differential rotation, and that the
axisymmetric component of the field should generally increase with rotation
rate.


\begin{theacknowledgments}
It is a pleasure to thank Juri Toomre, A. Sacha Brun, Andrew West, Mark
Miesch, Benjamin Brown, and Nick Featherstone for helpful discussions
and/or comments on this manuscript.  We also gratefully acknowledge
Isabelle Baraffe and Gilles Chabrier for supplying the 1--D stellar model
used here for initial conditions.  This work was supported by an NSF
Astronomy \& Astrophysics postdoctoral fellowship (AST 05-02413).

\end{theacknowledgments}




\end{document}